\documentclass[osajnl,preprint,showpacs,superscriptaddress,12pt]{revtex4-1} 

\usepackage{amsmath,amssymb,graphicx}

\begin{document}

\title{An optical analog of quantum optomechanics}

\author{B. M. Rodr\'{\i}guez-Lara}
\affiliation{Instituto Nacional de Astrof\'{i}sica, \'{O}ptica y Electr\'{o}nica \\ Calle Luis Enrique Erro No. 1, Sta. Ma. Tonantzintla, Pue. CP 72840, M\'{e}xico}
\email{bmlara@inaoep.mx}

\author{H. M. Moya-Cessa}
\affiliation{Instituto Nacional de Astrof\'{i}sica, \'{O}ptica y Electr\'{o}nica \\ Calle Luis Enrique Erro No. 1, Sta. Ma. Tonantzintla, Pue. CP 72840, M\'{e}xico}
\email{bmlara@inaoep.mx}

\begin{abstract}
We present a two-dimensional array of nearest-neighbor coupled waveguides that is the optical analog of a quantum optomechanical system. 
We show that the quantum model predicts the appearance of effective column isolation, diagonal-coupling and other non-trivial couplings in the two-dimensional photonic lattice under a standard approximation from ion-trap cavity electrodynamics.
We provide an approximate impulse function for the case of effective column isolation and compare it with exact numerical propagation in the photonic lattice.
\end{abstract}


\maketitle

\section{Introduction} \label{sec:S1}

Optical analogs of quantum processes \cite{Longhi2009p243,Longhi2011p453} may become a valuable resource for the design of integrated optics devices \cite{ElGanainy2013p161105,RodriguezLara2014p013802}.
In this spirit, a zoo of optical analogs to quantum mechanical systems and the algebraic methods to solve them provide a valuable toolbox for optical designers.
Here we want to show that a quantum analogy may be helpful in the design of optical integrated circuits even when it fails to provide a closed-form analytic impulse function for its optical simulator.
For this reason, we will consider the two-dimensional array of couple photonic waveguides described by the differential set
\begin{eqnarray}
- i \partial_{z} \mathcal{E}_{j,k} &=& \left(  \delta k + \omega_{m} j \right) \mathcal{E}_{j,k} + \nonumber 
\\ && k g  \left( \sqrt{j+1} \mathcal{E}_{j+1,k} + \sqrt{j} \mathcal{E}_{j-1,k}  \right) + \nonumber \\
&& d  \left( \sqrt{k+1} \mathcal{E}_{j,k+1} + \sqrt{k} \mathcal{E}_{j,k-1}  \right). \label{eq:2DLatt} 
\end{eqnarray}
This photonic lattice is a two-dimensional semi-infinite array of waveguides, where the effective refractive index depends from both horizontal and vertical positions in the array, and the couplings between nearest-neighbor waveguides go as the square root of their horizontal or vertical position. 
It can be seen as a two-dimensional realization of a Glauber-Fock oscillator lattice \cite{PerezLeija2012p013848} with some modifications.
It is straightforward to show that this differential set can be cast into a Schr\"odinger equation, $i \partial_{t} \vert \psi \rangle = \hat{H} \vert \psi \rangle$, with effective Hamiltonian, 
\begin{eqnarray}
\hat{H} =  \delta \hat{a}^{\dagger} \hat{a} + \omega_{m} \hat{b}^{\dagger} \hat{b} + g \hat{a}^{\dagger} \hat{a} \left( \hat{b}^{\dagger} + \hat{b} \right) + d \left( \hat{a}^{\dagger} + \hat{a} \right), \label{eq:OptoMech}
\end{eqnarray}
once we propose the variable change $z = -t$ and a wavefunction decomposition $\vert \psi \rangle = \sum_{n,m} \mathcal{E}_{m,n} \vert m \rangle_{b} \vert n \rangle_{a}$ where the state $\vert p \rangle_{q}$ is the $p$th Fock state of the $q$ oscillator, and the field at the $(j,k)$th waveguide is $\mathcal{E}_{j,k}$ with $j,k=0,1,2, \ldots$ and $ \mathcal{E}_{- \vert j \vert, -\vert k \vert} = 0$.  
The Hamiltonian in (\ref{eq:OptoMech}) describes an optomechanical model composed by a driven cavity coupled to a mechanical oscillator \cite{Pace1993p3173}. 
Here, the detunning between the cavity field mode described by the creation (annihilation) operators, $\hat{a}^{\dagger}$ ($\hat{a}$), and the pump field frequencies is given by $\delta = \omega_{f} - \omega_{p}$, the frequency of the mechanical oscillator described by the creation (annihilation) operators, $\hat{b}^{\dagger}$ ($\hat{b}$) is $\omega_{m}$, and the parameters $d$ and $g$  are the strenght of the pump field and the linear coupling between the field and the mechanical oscillator, respectively.
Up to our knowledge, it is not possible to provide a closed form analytic result for these equivalent models but some approximations can be made in the quantum model.
We will use this insight to predict the behavior of classical light propagating in the photonic lattices.

\section{An algebraic approach to Glauber-Fock oscillators.} \label{sec:S2}

\begin{figure}[t]
\centerline{\includegraphics[scale=1]{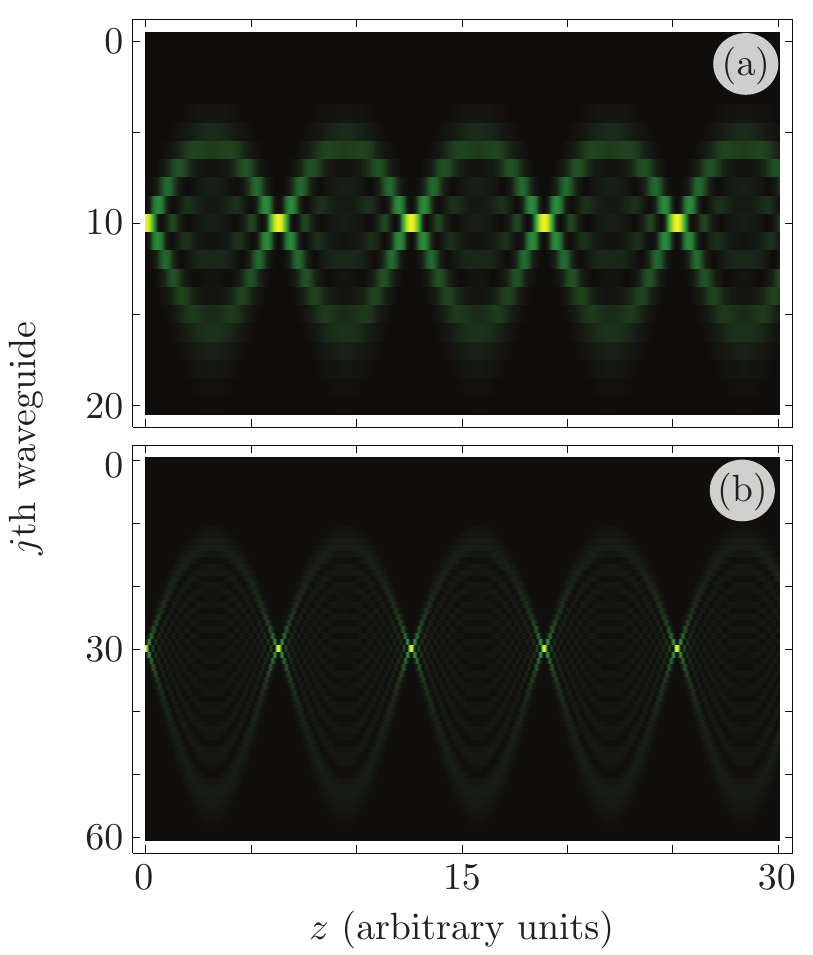}}
\caption{Evolution of the field intensity for an initial field impinging the (a)tenth and (b)thirtieth waveguide of a photonic lattice composed of 200 waveguides and described by Eq.(\ref{Eq:OscLatt}) with parameter set $ $ equal to (a) $ n g =   0.5 ~\omega_{m}$ and (b) $ n g = \omega_{m}$.} \label{fig:Fig1}
\end{figure}

The trivial case corresponds to the absence of driving in the optomechanical model, $d=0$,  
\begin{eqnarray}
\hat{H} =  \omega_{m} \hat{b}^{\dagger} \hat{b} + g \hat{a}^{\dagger} \hat{a} \left( \hat{b}^{\dagger} + \hat{b} \right), \label{eq:EffHam1}
\end{eqnarray} 
which has been shown to produce non-classical light states \cite{Mancini1997p3042,Bose1997p4175}, and its optical analogue is a one-dimensional Glauber-Fock oscillator lattice \cite{PerezLeija2012p013848},
\begin{eqnarray}
- i \partial_{z} \mathcal{E}_{j,n} = \omega_{m} j \mathcal{E}_{j,n} + n g  \left( \sqrt{j+1} \mathcal{E}_{j+1,n} + \sqrt{j} \mathcal{E}_{j-1,n}  \right). \nonumber \\ \label{Eq:OscLatt}
\end{eqnarray}
It is straightforward to diagonalize (\ref{eq:EffHam1}) via the coherent displacement $\hat{D}(g \hat{a}^{\dagger} \hat{a}/\omega_{m}) = e^{  g \hat{a}^{\dagger} \hat{a} \left( \hat{b}^{\dagger} - \hat{b} \right)/ \omega_{m}}$ \cite{Xu2013p053849,Xu2013p063819}, such that in diagonal form the effective Hamiltonian is given by
\begin{eqnarray}
\hat{H}_{D} &=& \hat{D}( g \hat{a}^{\dagger} \hat{a}/\omega_{m}) \hat{H} \hat{D}(-g \hat{a}^{\dagger} \hat{a}/\omega_{m}), \\
&=& \omega_{m} \hat{b}^{\dagger} \hat{b} - \frac{g^{2}}{\omega_{m}} \left( \hat{a}^{\dagger} \hat{a} \right)^{2},
\end{eqnarray}
and leads to a spectrum for the photonic lattice in the form:
\begin{eqnarray}
\Omega_{j} = \omega_{m} j - \frac{\left(g n\right)^{2}}{\omega_{m}}. 
\end{eqnarray}
The impulse function giving the field amplitude at the $p$th waveguide given that the initial field was located at the $q$th waveguide is 
\begin{eqnarray}
\mathcal{I}_{p,q}(z) &=& \sum_{j} \langle p \vert \hat{D}\left(-\frac{gn}{\omega_{m}}\right) \vert j \rangle \langle j \vert \hat{D}\left(\frac{gn}{\omega_{m}}\right) \vert q \rangle e^{i \Omega_{j} z
}, \label{eq:ImpFn1D}
\end{eqnarray}
and we can recover the terms given by
\begin{eqnarray}
\langle j \vert \hat{D}(\beta) \vert k\rangle &=& e^{-\frac{\vert \beta \vert^{2}}{2}} \sqrt{\frac{k!}{j!}} ~\beta^{j-k} L_{k}^{(j-k)} \left( \vert \beta \vert^{2} \right),\\ 
&=&  e^{-\frac{\vert \beta \vert^{2}}{2}} \sqrt{\frac{j!}{k!}} ~ (-\beta)^{k-j} L_{j}^{(k-j)} \left( \vert \beta \vert^{2} \right),
\end{eqnarray}
via reference \cite{Wunsche1991p359} where $L_{n}^{\beta}(x)$ stands for the generalized Laguerre polynomials \cite{Ederlyi1953}.
It have been shown that the Glauber-Fock oscillator is equivalent to a harmonic oscillator of constant mass and frequency where, for a fixed effective refractive index, the only way to change the oscillator frequency is the inclusion of second-neighbor couplings \cite{RodriguezLara2014p2083}. 
Thus, in our case a single-waveguide input will produce coherent oscillations and its spatial frequency will remain the same for a fixed effective refractive index, $\omega_{m}$, and variable coupling, $ng$, as shown in Fig. \ref{fig:Fig1}.

\section{The two-dimensional waveguide lattice.} \label{sec:S3}

\begin{figure}[t]
\centerline{\includegraphics[scale=1]{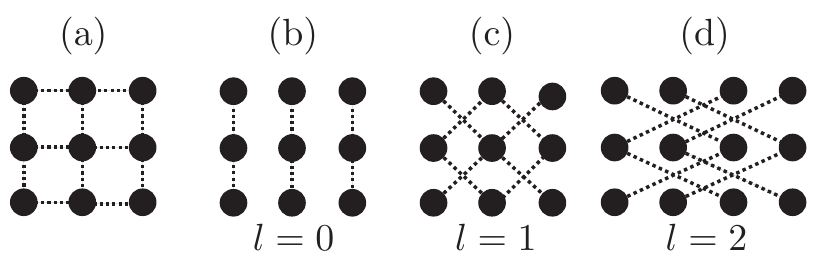}}
\caption {A schematic showing the (a) nearest-neighbor couplings in the optical lattice described by (\ref{eq:2DLatt}) and the effective couplings derived from the quantum analog for parameters $d \ll \omega_{m}$ and $\delta = l \omega_{m}$ with (b) $l=0$, (c) $l=1$ and (d) $l=2$. }  \label{fig:Fig2}
\end{figure}

So, we may expect some kind of oscillation in the two-dimensional model.
In order to gain some intuition let us cast the action of the coherent displacement defined before over the full Hamiltonian (\ref{eq:OptoMech}) \cite{Xu2013p053849,Xu2013p063819}, 
\begin{eqnarray}
\hat{H}_{D} &=&  \delta \hat{a}^{\dagger} \hat{a} - \frac{g^{2}}{\omega_{m}} \left( \hat{a}^{\dagger} \hat{a}  \right)^2 + \omega_{m} \hat{b}^{\dagger} \hat{b} + \nonumber \\
&& d  \left[ \hat{a}^{\dagger} e^{ \frac{g}{\omega_{m}} \left( \hat{b}^{\dagger} - \hat{b} \right) } + \hat{a} e^{- \frac{g}{\omega_{m}} \left( \hat{b}^{\dagger} - \hat{b} \right) } \right] .
\end{eqnarray}
At this point, we can use the time-dependent rotation $\hat{U}_{0}(t) = e^{-i t \left(  \delta \hat{a}^{\dagger} \hat{a} + \omega_{m} \hat{b}^{\dagger} \hat{b} \right) }$ and the new formulation of the Schr\"odinger equation has an effective Hamiltonian given by
\begin{eqnarray}
\hat{H}_{U} &=&  -\frac{g^2}{\omega_{m}} \left(\hat{a}^{\dagger} \hat{a} \right)^2 +  d \left[ \hat{a}^{\dagger} e^{i \delta t}  e^{ \beta \left( \hat{b}^{\dagger} e^{i \omega t} - \hat{b}^{\dagger} e^{-i \omega t}  \right) }  + \right. \nonumber \\
&& \left. \hat{a} e^{-i \delta t}  e^{ -\beta \left( \hat{b}^{\dagger} e^{i \omega t} - \hat{b}^{\dagger} e^{-i \omega t}  \right) }  \right], \\
&=& -\frac{g^2}{\omega_{m}} \left(\hat{a}^{\dagger} \hat{a} \right)^2 +  d \left[ \hat{a}^{\dagger} e^{i \delta t} D\left( \beta e^{i \omega t} \right)  +  \right. \nonumber \\
&& \left.   \hat{a} e^{-i \delta t} D\left( -\beta e^{i \omega t} \right)   \right], 
\end{eqnarray}
with $\beta = g / \omega_{m}$.
The second term of this Hamiltonian is equivalent to that of a trapped-ion setup \cite{MoyaCessa2012p229} and, as long as $d \ll \omega_{m}$, we can use the approach used in ion-trap quantum electrodynamics to approximate it to different types of coupling by setting the pump-cavity field detunning to $\delta = l \omega_{m} $ with $l=0,\pm1,\pm2,\ldots$ 
For positive values of $l$, we obtain
\begin{eqnarray}
\hat{H}_{U+} &\approx& -\frac{g^2}{\omega_{m}} \left( \hat{a}^{\dagger} \hat{a} \right)^2 + \nonumber \\
&&  d  e^{ - \frac{\vert \beta \vert^{2}}{2}}  \left\{  \hat{a}^{\dagger}  \frac{ \left(\hat{b}^{\dagger} \hat{b} \right)!}{\left(\hat{b}^{\dagger} \hat{b} + l \right)!} L_{\hat{b}^{\dagger} \hat{b}}^{(l)}\left( \beta^{2} \right) \left( - \beta \hat{b} \right)^{l} + \right. \nonumber \\
&& \left.  \hat{a} \left( - \beta \hat{b}^{\dagger} \right)^{l} \frac{ \left(\hat{b}^{\dagger} \hat{b} \right)!}{\left(\hat{b}^{\dagger} \hat{b} + l \right)!} L_{\hat{b}^{\dagger} \hat{b}}^{(l)}\left( \beta^{2} \right)  \right\}, \nonumber \\
\end{eqnarray}
and, as far as we know, it is not possible to create an optical analog involving photonic lattices for negative values of $l$.
Then, under the condition $d \ll \omega_{m}$, if we choose the lattice to show $\delta = 0$, the horizontal and vertical modes of the lattice should remain effectively uncoupled, as strange as it may sound. 
If we choose a paremeter value  $\delta = \omega_{m}$, then we have:
\begin{eqnarray}
\hat{H}_{U+} &\approx& \frac{g^2}{\omega_{m}} \left( \hat{a}^{\dagger} \hat{a} \right)^2 -  d \beta  e^{ - \frac{\vert \beta \vert^{2}}{2}}  \left( \hat{a} \hat{b}^{\dagger} + \hat{a}^{\dagger}  \hat{b}  \right), 
\end{eqnarray}
which tells us that horizontal and vertical modes interact such that $\mathcal{E}_{j,k}$ will effectively couple to $\mathcal{E}_{j\mp1,k\pm1}$. 
If we choose a  $\delta = 2 \omega_{m}$, then the effective Hamiltonian is
\begin{eqnarray}
\hat{H}_{U+} &\approx& \frac{g^2}{\omega_{m}} \left( \hat{a}^{\dagger} \hat{a} \right)^2 -  d \beta  e^{ - \frac{\vert \beta \vert^{2}}{2}}  \left( \hat{a} \hat{b}^{\dagger 2} + \hat{a}^{\dagger}  \hat{b}^{2}  \right), 
\end{eqnarray}
and the nearest-neighbor coupling lattice will behave as if second nearest diagonal neighbors were coupled; i.e., $\mathcal{E}_{j,k}$ will effectively couple to $\mathcal{E}_{j\mp1,k\pm2}$. 
Figure \ref{fig:Fig2} shows a diagram exemplifying these cases.
Note that by choosing $g \ll \omega_{m}$ and staying close to the first waveguides we could neglect the importance of the quadratic term $\left( \hat{a}^{\dagger} \hat{a} \right)^{2}$ and that experimental considerations should be taken into account for the  definition of the parameter ranges as stronger couplings for waveguides far from the zeroth waveguide will induce second- and higher-neighbor couplings.

If we were able to diagonalize $\hat{H}_{U}$, then the impulse function giving the field amplitude at the $(p,q)$th waveguide for an input at the $(r,s)$th waveguide would be given by
\begin{eqnarray}
\mathcal{I}_{(p,q),(r,s)}&=& \sum_{j=0}^{\infty} e^{ i z \left( \delta p + \omega_{m} j\right)} \langle q \vert D\left( - \frac{g p}{\omega_{m}} \right) \vert j \rangle \times \nonumber \\
&& \langle p, j \vert e^{i z \hat{H}_{U}}  D\left( \frac{g r}{\omega_{m}} \right) \vert r,s \rangle.
\end{eqnarray}
As we mentioned before, we are not able to do so but we can provide certain approximations.
For example, the quantum optomechanical model tells us that if we consider $\delta= 0$ then the effective Hamiltonian does not involve any term with the form $\left( \hat{a}^{\dagger} + \hat{a} \right)$ that relates to coupling between horizontal waveguides in the photonic lattice. 
Then, we can argue that we will effectively have a series of vertical Glauber-Fock oscillators which do not see each other. 
This is confirmed by the approximate impulse function,
\begin{eqnarray}
\mathcal{I}_{(p,q),(r,s)}&\approx& \left\{ \begin{array}{ll}
0, & p\ne r, \\
e^{ i z \left( \delta p - \frac{g^{2}p^{2}}{\omega_{m}}\right)} \sum_{j=0}^{\infty} e^{ i z \omega_{m} j} \times & \\
\langle q \vert D\left( - \frac{g p}{\omega_{m}} \right) \vert j \rangle \langle j \vert D\left( \frac{g p}{\omega_{m}} \right) \vert p \rangle, & p=r .
\end{array} \right. \nonumber \\
\end{eqnarray}
This impulse function is just the impulse function of a Glauber-Fock oscillator in  (\ref{eq:ImpFn1D}) times a phase factor. 
We already know from the harmonic oscillator analog \cite{RodriguezLara2014p2083} and Fig. \ref{fig:Fig1} that the oscillation frequency between different single-waveguide inputs is the same. 
Thus, if we introduce a horizontal light line, the propagating field in each and every excited waveguide should oscillate and come back to the original state almost at the same time.
\begin{figure}[b]
\centerline{\includegraphics[scale=1]{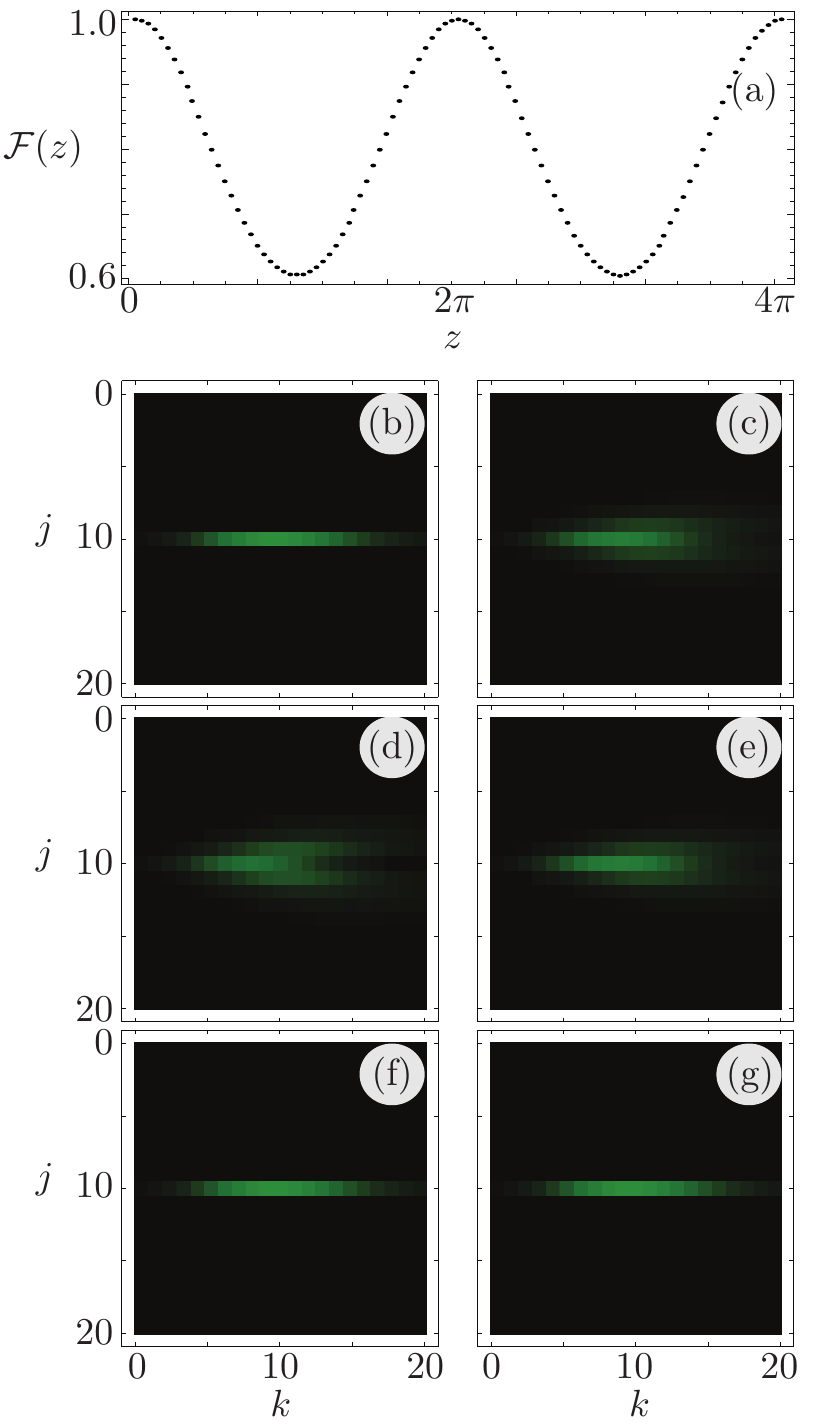}}
\caption { (a) The fidelity of the propagating field with respect to the input field. Numerical propagation of the field intensity at the two-dimensional lattice described by Eq.(\ref{eq:2DLatt}) with parameter set $\left\{ \delta, g, d \right\} = \left\{ 0, 0.01, 0.01 \right\} \omega_{m} $ at distances (b) $z=0$, (c) $z= \pi/2$, (d)  $z= \pi$, (e)  $z= 3 \pi/2 $, (f) $z=2 \pi$, (g) $z = 4 \pi$. }  \label{fig:Fig3}
\end{figure}
Figure \ref{fig:Fig3} shows the numerical propagation in a finite lattice composed by  ten thousand waveguides in a $100 \times 100$ array described by (\ref{eq:2DLatt}).
The lattice size is adequate to keep the propagating field away from the end boundary and the parameters are chosen to fulfill $\delta=0$, $d=g= 0.01~\omega_{m}$.
The initial field distribution $\mathcal{E}_{j,k} = \delta_{j,10} ~ e^{-\vert \alpha \vert^{2}/2} \alpha^{k} (k!)^{-1/2} $ corresponds to a quantum analog where the mechanical oscillator is in the tenth Fock state and the field oscillator in a coherent state, $ \vert 10, \alpha \rangle$,  with coherent parameter $\alpha = \sqrt{10}$.
Figure \ref{fig:Fig3}(a) shows the fidelity between the original input field and the propagating field, $\mathcal{F}(z)= \vert \mathbf{E}(0)\cdot \mathbf{E}(z) \vert$, where the field amplitudes vector is defined as $\mathbf{E} = \left\{ \mathcal{E}_{0,0},\mathcal{E}_{0,1}, \ldots, \mathcal{E}_{1,0}, \mathcal{E}_{1,1}, \ldots \right\}$.
We can see that the field performs a periodical oscillation as expected from the approximate impulse function.
Figures \ref{fig:Fig3}(b) to Fig. \ref{fig:Fig3}(g) show the field intensity in the two-dimensional lattice at positions $z=0$, Fig. \ref{fig:Fig3}(b), $z= \pi/2$, Fig. \ref{fig:Fig3}(c), $z= \pi$, Fig. \ref{fig:Fig3}(d), $z= 3 \pi/2 $, Fig. \ref{fig:Fig3}(e), $z=2 \pi$, Fig. \ref{fig:Fig3}(f), and $z = 4 \pi$, Fig. \ref{fig:Fig3}(g).  
Of course, this is only an approximation and the effective isolation between waveguide columns is not perfect as Fig. \ref{fig:Fig4} shows for an initial field distribution $\mathcal{E}_{j,k} = \delta_{k,10}  e^{-\vert \alpha \vert^{2}/2} \alpha^{j} (j!)^{-1/2} $ that corresponds to the mechanical oscillator in a coherent state and the field in a Fock state, $ \vert  \alpha, 10 \rangle$ with coherent parameter $\alpha = \sqrt{10}$, and looks like a vertical light line. 

\begin{figure}[t]
\centerline{\includegraphics[scale=1]{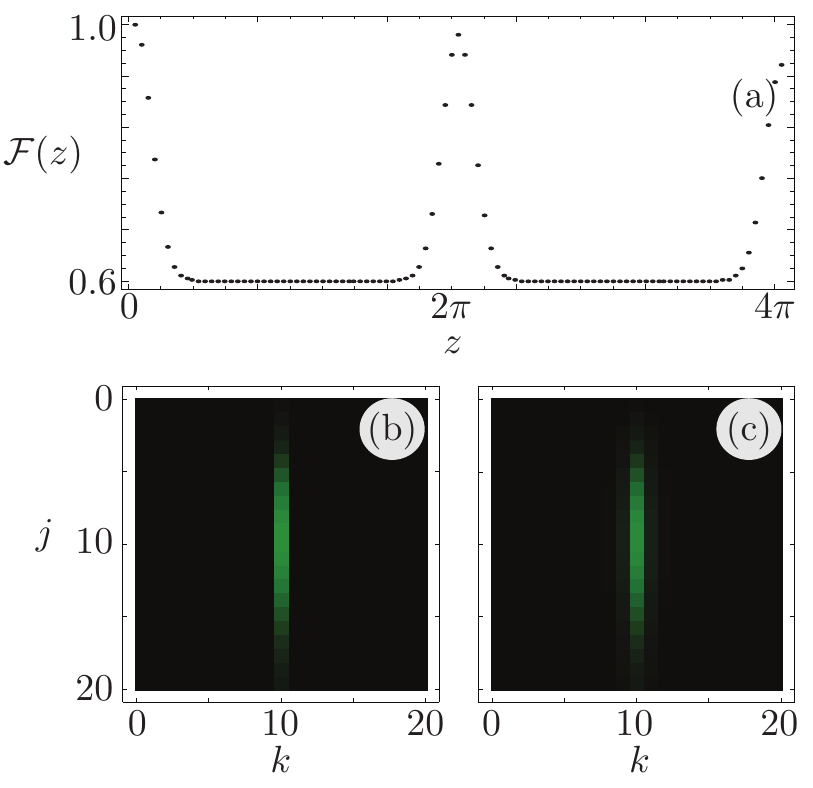}}
\caption { (a) The fidelity of the propagating field with respect to the input field. Numerical propagation of the field intensity at the two-dimensional lattice described by Eq.(\ref{eq:2DLatt}) with parameter set $\left\{ \delta, g, d \right\} = \left\{ 0, 0.01, 0.01 \right\} \omega_{m} $ at distances (b) $z=0$, (c) $z=2 \pi$. }  \label{fig:Fig4}
\end{figure}

\section{Conclusions} \label{sec:S4}

In summary, we have shown that analogies between optical and quantum systems may help the optical designer even in those cases where an exact solution cannot be procured.
We chose as an example a two-dimensional array of waveguides that is the optical analog of a quantum optomechanical system.
First, we show that the closed quantum model is related to the one-dimensional Glauber-Fock oscillator and calculate its spectra and impulse function using an algebraic approach that some may find simpler than the Bargmann formalism used in the literature.
Then, we move to the driven quantum model and bring forward a method used in trapped-ion cavity quantum electrodynamics to approximate the dynamics in a given regime.
Such an approach leads us to approximate dynamics where different effective couplings between waveguides appear; e.g., uncoupling between columns or effective diagonal coupling.
We chose to work with the case where waveguide columns do not effectively couple, which means that each and every column of the array behaves as an independent Glauber-Fock oscillator.
We provide an approximate impulse function and a numerical exact propagation in the original two-dimensional lattice that supports our prediction.




\begin{thebibliography}{10}
\newcommand{\enquote}[1]{``#1''}

\bibitem{Longhi2009p243}
S.~Longhi, \enquote{Quantum-optical analogies using photonic structures,} Laser
  Photon. Rev. \textbf{3}, 243 -- 261 (2009).

\bibitem{Longhi2011p453}
S.~Longhi, \enquote{Classical simulation of relativistic quantum mechanics in
  periodic optical structures,} Appl. Phys. B \textbf{104}, 453 -- 468 (2011).

\bibitem{ElGanainy2013p161105}
R.~El-Ganainy, A.~Eisfeld, M.~Levy, and D.~N. Christodoulides, \enquote{On-chip
  non-reciprocal optical devices based on quantum inspired photonic lattices,}
  Appl. Phys. Lett. \textbf{103}, 161105 (2013).

\bibitem{RodriguezLara2014p013802}
B.~M. Rodr{\'\i}guez-Lara, H.~M. Moya-Cessa, and D.~N. Christodoulides,
  \enquote{Propagation and perfect transmission in three-waveguide axially
  varying couplers,} Phys. Rev. A \textbf{89}, 013802 (2014).

\bibitem{PerezLeija2012p013848}
A.~Perez-Leija, R.~Keil, A.~Szameit, A.~F. Abouraddy, H.~Moya-Cessa, and D.~N.
  Christodoulides, \enquote{Tailoring the correlation and anticorrelation
  behavior of path-entangled photons in glauber-fock lattices,} Phys. Rev. A
  \textbf{85}, 013848 (2012).

\bibitem{Pace1993p3173}
A.~F. Pace, M.~J. Collett, and D.~F. Walls, \enquote{Quantum limits in
  interferometric detection of gravitational radiation,} Phys. Rev. A
  \textbf{47}, 3173 -- 3189 (1993).

\bibitem{Bose1997p4175}
S.~Bose, K.~Jacobs, and P.~L. Knight, \enquote{Preparation of nonclassical
  states in cavities with a moving mirror,} Phys. Rev. A \textbf{56}, 4175 --
  4186 (1997).

\bibitem{Mancini1997p3042}
S.~Mancini, V.~I. Man'ko, and P.~Tombesi, \enquote{Ponderomotive control of
  quantum macroscopic coherence,} Phys. Rev. A \textbf{55}, 3042 -- 3050
  (1997).

\bibitem{Xu2013p053849}
G.-F. Xu and C.~K. Law, \enquote{Dark states of a moving mirror in the
  single-photon strong-coupling regime,} Phys. Rev. A \textbf{87}, 053849
  (2013).

\bibitem{Xu2013p063819}
X.-W. Xu, H.~Wang, J.~Zhang, and Y.-X. Liu, \enquote{Engineering of
  nonclassical motional states in optomechanical systems,} Phys. Rev. A
  \textbf{88}, 063819 (2013).

\bibitem{Wunsche1991p359}
A.~W{\"u}nsche, \enquote{Displaced fock states and their connection to
  quasiprobabilities,} Quantum Opt. \textbf{3}, 359 -- 383 (1991).

\bibitem{Ederlyi1953}
A.~Ed\'erlyi, \emph{Higher trascendental functions}, vol.~1 (McGraw-Hill,
  1953).

\bibitem{RodriguezLara2014p2083}
B.~M. Rodr{\'\i}guez-Lara, P.~Aleahmad, H.~M. Moya-Cessa, and D.~N.
  Christodoulides, \enquote{Ermakov-lewis symmetry in photonic lattices,} Opt.
  Lett. \textbf{39}, 2083--2085 (2014).

\bibitem{MoyaCessa2012p229}
H.~Moya-Cessa, F.~Soto-Eguibar, J.~M. Vargas-Mart\'{\i}nez, R.~Ju\'arez-Amaro,
  and A.~Z{\'u}{\~n}iga-Segundo, \enquote{Ion-laser interactions: The most
  complete solution,} Phys. Rep. \textbf{513}, 229 -- 261 (2012).

\end{thebibliography}

\end{document}